\documentclass[a4paper,10pt,twocolumn,twoside]{article}
\usepackage{amsmath, amsthm, amssymb}
\usepackage[english]{babel}
\usepackage{chicago}
\usepackage[pdftex]{graphicx}
\usepackage{hyperref}
\usepackage{algorithmic}
\usepackage{algorithm}
\usepackage[utf8]{inputenc}
\usepackage{geometry}
\hypersetup{
pdfauthor=Havemann \& Gl\"{a}ser \& Heinz,
pdftitle=Detecting Overlapping Link Communities,
colorlinks=true,
linkcolor=black,
anchorcolor=black,
citecolor=blue,
urlcolor=black,
menucolor=black
}

\title{Detecting Overlapping Link Communities by Finding \\Local Minima of a Cost Function
with a Memetic Algorithm\\Part 1: Problem and Method}

\author{Frank Havemann\footnote{Institut f\"{u}r Bibliotheks- und Informationswissen\-schaft, Humboldt-Universit\"at zu Berlin, D 10099 Berlin, Doro\-theen\-str. 26 (Germany)} 
\and
Jochen Gl\"{a}ser\footnote{Center for Technology and Society, 
TU Berlin (Germany)}
\and Michael Heinz\footnote{Institut f\"{u}r Bibliotheks- und Informationswissen\-schaft, Humboldt-Universit\"at zu Berlin}
}
\date{}
\begin{document}

\maketitle

\begin{abstract}
We propose an algorithm for detecting communities of links in networks which uses local information, is based on a new evaluation function, and
allows for pervasive overlaps of communities. The complexity of the clustering task requires the application of a memetic algorithm that combines probabilistic evolutionary strategies with deterministic local searches. 
In Part 2 we will present results of experiments with citation networks.
\end{abstract}

\section{Introduction}
\label{intro}
Communities in networks are commonly defined as  cohesive subgraphs which are well separated from the rest of the network. This vague concept of communities  is operationalised in a variety of ways  \cite{fortunato2010community}.
The utility of algorithms for the detection of communities in networks partly depends on their `conceptual fit', i.e.\ on the degree to which they match properties of the phenomenon that is represented 
\cite{hric_community_2014}. Achieving such a conceptual fit may require unusual combinations of ideas from network analysis, as is the case with the question and the algorithm presented in this paper. 

Consider the following three properties of a network and the task of community detection.  First, links between nodes contain better information about communities than the nodes that are to be clustered. In this case, link clustering appears to be the method of choice. Constructing communities by clustering links has been proposed by \citeN{evans2009line} and by \citeN{ahn2010link} as a method for 
the construction of overlapping communities of nodes. In addition, clustering links is likely to be advantageous whenever the information asymmetry described above occurs, i.e.\ whenever links rather than nodes have the real-world properties whose similarity shall be reflected by clusters.

Second, overlapping communities must be a possible outcome of the algorithm 
because the real-world phenomenon under investigation is known to have such a structure.
For the same reason, pervasive overlaps must be possible, i.e.\ overlaps that extend to all nodes rather than 
just the boundary nodes of a community. The construction of overlapping communities is by now a well-known and frequently addressed problem of network analysis \cite{fortunato2010community,Xie:2013:OCD,amelio_overlapping_2014}.

Third, 
the phenomena to be represented by communities are local in that they emerge from local interactions represented by neighbouring nodes and links in the network. If this 
is the case, the use of local rather than global information may return better communities and a better community structure of the network \cite{clauset2005flc,lancichinetti2009detecting,havemann2011identification}. 

All three ideas have been developed in network analysis. However, as reviews of algorithms indicate \shortcite{fortunato2010community,Xie:2013:OCD,amelio_overlapping_2014}, link clustering, pervasively overlapping communities and use of local information have not yet been combined all three, possibly because the task for which this is necessary has not yet arisen.\footnote{The only apparent exception is the work by Lei Pan \textit{et al.}\ which, however, compromises in two expects, global information used in the end and no pervasive overlap because link clusters are disjunct \cite{pan_detecting_2011,pan_link_2012}. Furthermore, they differ from our approach because they propose an evaluation function for link clustering which is derived within the node clustering approach.}

There is at least one task for which this combination of link-based approach, pervasive overlaps and local approach is necessary, namely the detection of thematic structures (topics) in networks of papers. 

In networks of papers and their cited sources, citation links (links between a publication and the sources it cites) are thematically more homogenous than nodes (papers), and thus provide better information for clustering, than the papers themselves. While papers commonly belong to more than one scientific topic, many citation links can be assumed to be homogenous in that the link between paper and source belongs to only one topic. If it belongs to more than one topic these topics often are not very distant from each other. 

Scientific topics are known to overlap pervasively, which means that their reconstruction as communities of papers must reflect this pervasive overlap. Topics are also locally emergent phenomena in that they represent coinciding and mutually referring perspectives of researchers (the authors of the papers). 

In order to reconstruct scientific topics from networks of papers and their cited sources, then, we need an algorithm that clusters links, can construct pervasively overlapping communities, and uses 
mainly
local information. In this paper, we present such an algorithm (in Part 1) and its application to 
citation networks 
(in Part 2).
We propose a local cost function for the independent evaluation of each link community by relating its external to its total connectivity in the network. The cost function is almost completely  based on local information, the only global information used is the number of links in the whole network. The independent evaluation of each subgraph with a local cost function means that communities can be constructed independently from each other, which enables pervasive overlaps. 

The cost function we propose for subgraph evaluation is solely based on the network's topology and not on  link similarity.
Generally, clustering by optimising a (global or local) evaluation function needs no measure of similarity of clustered elements but results in clusters the elements of which are seen as similar in some sense.  In contrast, the approach to link clustering proposed by \shortciteN{ahn2010link} is based on link similarity.  The authors estimate the similarity of two links by comparing their sets of neighbouring nodes. This is not very appropiate for citation links because we would estimate thematic similarity of thematically nearly homogenous elements (citation links) with sets of very inhomogenous elements (papers, cited sources). In the case of citation networks, 
it would be better to measure link similarity by
using textual information from citing and cited documents.

The local construction of topics, their varying size and pervasive overlaps make it likely that
topics form a poly-hierarchy 
i.e.\ a hierarchy where a smaller topic can be a subtopic of two or more larger topics that have no hierarchical subtopic relation. This poly-hierarchy of topics should be reflected in a poly-hierarchy of communities.

Communities without sub-communities can be well separated and very cohesive, too,  but inside larger communities there can exist well separated sub-communities which diminish the cohesion of their super-community.

Since the cost landscape of link communities has many local minima, purely deterministic search strategies are not efficient. This is why we designed 
a memetic search that combines an evolutionary algorithm with deterministic adjustments in the cost landscape. 
Evolutionary algorithms have already been used for identifying communities in networks \cite[p.\ 106]{fortunato2010community}.
Some authors have even applied evolutionary algorithms to link clustering but all used global evaluation functions \cite{pizzuti_overlapped_2009,li_discovering_2013,shi_link_2013}.
Memetic evolutionary algorithms have also been applied to reconstruct communities but only for node clustering and only with global evaluation functions \cite{gong_memetic_2011,pizzuti_boosting_2012,gach_memetic_2012,ma_multi-level_2014}.

\section{Strategy}

The strategy we apply in response to the three challenges described in the introduction consists of three main steps. We develop an evaluation function for link communities that uses local information. This evaluation function makes it possible to construct each community independently from all others, which in turn enables pervasive overlaps because inner links (links all of whose neighbours are community members) of one community can also be inner links of another community. We then design an algorithm that 
constructs local communities.

For the first step, we 
followed a suggestion by \citeN{evans2009line} to obtain link clusters by clustering vertices in a network's line graph.
We defined a local cost function $\Psi(L)$ in the line-graph approach which we call \textit{ratio node-cut}. 
It can be used to identify link communities 
by finding local minima in the cost landscape. 
Since  $\Psi(L)$ evaluates the boundary between a subgraph and the 
rest of the
network, communities can be constructed independently of all other communities. 

The cost landscape of $\Psi(L)$ is often very rough i.e.\ has many local minima that may correspond to very similar subgraphs. Therefore, the resolution of the algorithm must be defined by setting a minimum distance (number of links that differ) between subgraphs corresponding to different local minima. We define the range of a community as a distance in which no subgraph exists that has a lower $\Psi$-value. 

Since the task of finding communities in large networks is always very complex, heuristics must be applied. This applies even more strongly to link clustering because networks contain many more links than nodes, and particularly to the rough $\Psi$-landscape. We chose an evolutionary algorithm but accelerate evolution by combining it with a deterministic local search in the cost landscape. This approach is called memetic \cite{neri2012handbook}. Memetic algorithms can also find local optima of a local cost function \cite{vitela_sequential_2012}.

In evolutionary algorithms, individuals occupy places in the cost (or fitness) landscape. In our local algorithm, populations are sets of different subgraphs. We start with a random initialisation of the population of some definite size. The genetic operators of crossover, mutation, and selection are repeatedly applied to move the population into optima. In memetic algorithms each crossover and each mutation is followed by a local search. 

In large networks exploring the cost landscape by adding or removing individual links is very time-consuming. We therefore begin the search with a coarse search phase that adds or removes groups of links by adding or removing nodes with all their links, and follow it with fine search phase, namely 
link-wise memetic evolution or at least
 a link-wise local search.
\label{node-wise}

\section[The cost function:\textit{ratio node-cut}]{The cost function:\\\textit{ratio node-cut}} 
\subsection[Node-induced and link-induced subgraphs]{Node-induced and\\link-induced subgraphs}

Traditionally, the boundary of a community is drawn between nodes and therefore cuts the links between nodes inside and outside the community.  
If we consider communities as clusters of links rather than nodes, the perspective must be reversed. While the boundary of a node community cuts links, the boundary of a link community cuts nodes.

A node community is a connected subgraph defined by a node set $C$. It contains all links existing between nodes in  $C$. A link community is a connected link-induced subgraph.  It 
contains all nodes 
attached to links of a given set $L$. There can be links existing between a link community's nodes which are not in $L$.

Cost functions of a subgraph can be defined by relating a measure of external to a measure of total connectivity. 
This ratio should be minimal for well separated and cohesive subgraphs i.e.\ for communities.

Node communities can be defined as connected subgraphs corresponding to minima in cost landscapes where places correspond to node-induced subgraphs. 
Correspondingly, link communities can be defined 
as connected subgraphs corresponding to minima in cost landscapes where places correspond to link-induced subgraphs.

In the following, we only consider connected unweighted graphs $G = (V,E)$.
The number of edges (or links) is $m = | E |$,  the number of vertices (or nodes) is $n = | V |$. With $k_i$ we denote the degree of node $i$. The internal degree of node $i$, denoted by $k_i^\mathrm{in}(L)$, is the number of links attached to node $i$ which are in link set $L$. The external degree of node $i$ is $k_i^\mathrm{out}(L) = k_i - k_i^\mathrm{in}(L)$.

\subsection{External connectivity}

We first consider measures of \textit{external connectivity} of a subgraph which are useful for constructing node or link communities. The simplest measure of external connectivity of a node-induced subgraph is the \textit{cut size} that equals the sum of weights of boundary links i.e.\ the links connecting the subgraph with the rest of the graph \cite[p.\ 92]{fortunato2010community}. 
If link weights represent electrical conductance, cut size measures the total conductance of all boundary links.  
Cut size can be calculated as the sum of external degrees $k_i^\mathrm{out}(L)$ of boundary nodes (subgraph members with boundary links). 

Applying these considerations to the external connectivity of a link-induced subgraph leads to a simple measure of external connectivity as the 
sum of $k_i^\mathrm{out}k_i^\mathrm{in}/k_i$ of boundary nodes: 
\begin{equation}
\sigma(L) = \sum_{i = 1}^n \dfrac{k_i^\mathrm{out}(L)k_i^\mathrm{in}(L)}{k_i}.
\end{equation} 
Only for boundary nodes of $L$ we have $k_i^\mathrm{out}k_i^\mathrm{in} > 0$. That means, we can restrict the sum in the  formula to boundary nodes.
In function $\sigma(L)$ the external degrees $k_i^\mathrm{out}$ are weighted with subgraph membership-grade $k_i^\mathrm{in}/k_i$ of the boundary nodes.
The function $\sigma(L)$ can be derived from the total conductance or cut size of link sets in the graph's line graph if the line graph's edges are weighted with $1/k_i$---a weighting proposed by \citeN{evans2009line}.
The derivation can be found in Appendix \ref{app:sigma}.

Each term of $\sigma(L)$ equals the conductance of a boundary node $i$ i.e.\ the total conductance for currents flowing out of the 
subgraph
through this node. 
We call 
$\sigma(L)$ the \textit{node cut} of a link-induced subgraph. 

\subsection[Internal and total connectivity]{Internal and total connectivity}

Now we discuss measures of \textit{internal and total connectivity} of subgraphs induced by node and by link sets, respectively.
In the case of node-induced subgraphs $k_\mathrm{in}(C)=\sum_{i \in C} k^\mathrm{in}_i(C)$ is an appropriate measure of internal connectivity of node set $C$.
Total connectivity of $C$ is then 
the sum of degrees of all nodes in $C$:
\begin{equation}
k_\mathrm{total}(C)=\sum_{i \in C} k^\mathrm{in}_i(C) + k^\mathrm{out}_i(C)=\sum_{i \in C} k_i.  \end{equation} 
For  a link-induced subgraph we can use the sum of 
internal degrees, weighted with their membership, as a measure of internal connectivity: 

\begin{equation}
\tau(L) = \sum_{i = 1}^n \dfrac{k_i^\mathrm{in}(L)k_i^\mathrm{in}(L)}{k_i}.
\end{equation} 
The sum is restricted to nodes attached to links in $L$ because other nodes have $k_i^\mathrm{in}(L)=0$. Total connectivity of $L$ is then given by the sum $\sigma(L)+\tau(L)= \sum_{i = 1}^nk_i^\mathrm{in}(L)=k_\mathrm{in}(L)=2|L|$.  The derivation can be found in Appendix \ref{app:sigma}.

\subsection{Cost function}
Relating external to total connectivity leads us to cost functions 
whose minima correspond to 
well separated and cohesive 
subgraphs.
On the other hand, we also achieve a size normalisation when we divide external by total connectivity.
This is welcome, because the boundary length (measured by external connectivity) tends to increase with size (here measured by total connectivity 
$k_\mathrm{in}(L) = 2|L|$%
)---at least for not too large 
subgraphs 
in not too small networks. If a 
subgraph 
occupies more than one half of the network its boundary tends to become shorter with increasing size. A simple size normalisation that accounts for the finite size of the network is achieved by adding to the external-total ratio of a 
subgraph 
the same ratio of its complement. For small 
subgraphs  
in a large network the second ratio is very small. For node%
-induced subgraphs 
this normalisation was introduced by \citeN{wei_towards_1989} and named \textit{ratio cut}. For link-induced subgraphs 
we analogously define a cost function \textit{ratio node-cut} as 
\begin{eqnarray}
\Psi(L)&=&\dfrac{\sigma(L)}{k_\mathrm{in}(L)}+\dfrac{\sigma(E\backslash L)}{k_\mathrm{in}(E\backslash L)}
\\
&=& \dfrac{\sigma(L)}{k_\mathrm{in}(L)(1-k_\mathrm{in}(L)/2m)}.
\end{eqnarray} 
The expression on the r.h.s. is obtained because $\sigma(E\backslash L)=\sigma(L)$ and $k_\mathrm{in}(E\backslash L)=2m-k_\mathrm{in}(L)$.
Ratio node-cut $\Psi$ is not strictly local but the only global information needed here is the total number of links $m$.
In the limit of small subgraphs in large networks we achieve approximately strict locality because we have $k_\mathrm{in}(L) \ll 2m$ and we therefore obtain  
\begin{equation}
\Psi(L) \approx \dfrac{\sigma(L)}{k_\mathrm{in}(L)},%
\end{equation} 
which equals a strictly local cost function 
for the construction 
of link communities introduced by us earlier \shortcite{havemann2012evaluating}.
Our cost function $\Psi(L)$ rewards separation of link community $L$ but not really its cohesion. \citeN{yang2012defining} found that  
evaluating node subgraphs with \textit{conductance}---a measure analogue to $\sigma(L)/k_\mathrm{in}(L)$ in the world of node communities---can also lead to communities with low cohesion.

That means, using cost function $\Psi$ we 
emphasize separation and require only a minimal cohesion of subgraphs, which is expressed by the demand that subgraphs must be connected.
Otherwise an unconnected subgraph with parts in very different regions of the network could be a community.

Function $\sigma(L)$ vanishes for the empty subgraph with $L = \emptyset$ and for the full graph with $L = E$. In both cases, the denominator of the cost function also vanishes and we obtain zero divided by zero but it makes sense to define $\Psi(E)=\Psi(\emptyset)=1$
because $\Psi$ of one link (1, 2) with vanishing weight $w_{12}$  approximates 1: 
\begin{equation}
\Psi((1, 2)) =w_{12}\dfrac{(k_1-w_{12})/k_1 + (k_2-w_{12})/k_2}{2w_{12}(1-2w_{12}/2m)}.
\end{equation} 
Our cost function is symmetric: $\Psi(L)=\Psi(E\backslash L)$, i.e.\ the cost function is the same for a link-induced subgraph and the subgraph induced by the complementary link set $E\backslash L$.

\subsection{The cost landscape}
Each place in the cost landscape represents a link-induced subgraph.
Two places in the landscape have a direct relation if and only if the corresponding subgraphs differ in one link.
The height of each place is given by the value of the cost function $\Psi(L)$. 
The global minimum of the cost function is reached for a division of the set $E$ of all links that produces the two best link communities in terms of
separation. 
As a simple example, we determined the $\Psi$-landscape of the bow-tie graph (Figure  \ref{bow-tie}, for calculations see Appendix \ref{sec:bowtie}). We expect a cut through the central node to be the best division in two link communities (the two triangles). Indeed, the landscape has two minima with 
$\Psi = 1 / 3$, which correspond to the two triangles. There are no further local minima.

 \begin{figure}[!b] 
\begin{center}  \includegraphics[width=2in]{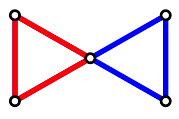} 
\end{center} 
\caption{Bow-tie graph} \label{bow-tie} 
\end{figure}%

We do not restrict the search for link communities to finding only the global minimum but
define a link community as a connected link-induced subgraph which corresponds to any local minimum in the $\Psi$-landscape. 
Since the $\Psi$-landscape of larger graphs contains many local minima, we need a filter to select the locally 
best link communities. For this reason, we restrict our search to those minima with a sufficiently large distance to any lower place in the cost landscape. Thus, we have to define the resolution of the search by defining this minimal distance in the landscape.
The appropriate resolution depends on the research question about the phenomenon represented by the network. The extent to which two communities
should differ in content (of links) to consider them as different depends on the question asked about communities. 

Another place in the cost landscape is reached by adding links to and by removing links from the subgraph corresponding to the starting place.
The distance between two places in the cost landscape equals the sum of the number of links we have to add and to exclude.
In other words: the distance is the size of the symmetric difference between the two link sets. 
We define the range of a community as the minimal distance to a subgraph with lower cost.
Within a community's range there is no better subgraph. 
The resolution of 
a search for communities can be defined as the minimal range of communities that are
accepted as valid 
solutions. 
Depending on the networks real background, a relative resolution can be more appropriate.
That means, we demand that any valid 
community should have a range which is larger than a certain percentage of its size.

In order to determine the range of a community we would need to know its whole environment up to the distance to the nearest lower place in the cost landscape. Otherwise, a lower place only determines an upper bound of the community's range. However, searching the whole environment of a subgraph is practically impossible for large networks. A selective search is necessary, which is why we apply evolutionary and deterministic greedy algorithms. If these algorithms find an upper bound smaller than the set resolution, we can deselect the community. If they don't, the community is provisionally kept but can later be replaced by a better community within its minimal range defined by the set resolution. We assume, however, that later found better solutions  differ only in some links.

\section{Memetic search}

Memetic algorithms combine random evolution with deterministic local search.
In this section, we describe 
 \begin{enumerate}
\item the local search we apply, called adaptation for short,
  \item  our implementation  of the evolutionary approach, 
 \item  the genetic operators of mutation, cross\-over, and selection  we employ in the evolutionary approach.
  \end{enumerate}

The memetic algorithm is applied in the search for link communities, which can be done by exploring the cost landscape of a  
network by adding or removing individual links. For large subgraphs this is very time-consuming. We therefore split the search in a coarse phase, in which we add or remove nodes with all their links to other nodes in the subgraph, and a finer link-wise 
search, which is applied after communities have been identified by a node-wise search. 
After communities with a minimal range defined by the set resolution are found in a node-wise memetic search, they are subjected to a link-wise memetic search or at least a link-wise local search.

\subsection{Local search}
The local search in the cost landscape applies a greedy algorithm for finding local cost minima that correspond to communities. The algorithm starts from the place occupied by the current subgraph and moves to subgraphs with lower $\Psi$-values. The algorithm is greedy because it always chooses the step that brings the biggest decrease or the smallest increase  of $\Psi$. A step includes or excludes a node with all their links to the nodes already in a subgraph in node-wise local search, 
and includes or excludes an individual link in link-wise local search.

A valid community can be made invalid and replaced by a better one if the better one is within its minimal range which is set by the resolution parameter. 
Therefore, the local search 
has not to find subgraphs with lower cost in each step but 
can go a number of steps by `tunneling' through `barriers' in the landscape (areas with higher $\Psi$) before reaching lower values which 
invalidate the community at the tunnel's entrance. Tunneling 
makes the algorithm more efficient. The maximum length of a tunnel through a barrier of higher $\Psi$-values is determined by the set resolution.

The local search can begin by a series of either inclusions or exclusions of nodes (links). When no further improvement can be achieved, the search switches from inclusion to exclusion or vice versa. Inclusion and exclusion are continued until no further improvement is possible.

If the exclusion of nodes 
fragments a subgraph, we proceed with the subgraph's main component. 
In the link-wise local search the greedy algorithm is allowed to go through intermediary states representing unconnected subgraphs. 
At the end of the link-wise local search we determine all components of the subgraph. If the subgraph is unconnected we repeat the procedure for each component until we obtain only connected subgraphs with minimal cost.

A greedy algorithm is efficient because the cost reduction for all possible cases of including a neighbour must be calculated only at the beginning of the local search. In the subsequent steps, we only calculate or recalculate cost reductions achieved by adding neighbours of the link (or node) included. 
Analogously, we proceed when excluding boundary nodes or links \shortcite[Appendix]{havemann2012evaluating}. Otherwise it would be more efficient to include or exclude just the first node (link) which reduces cost.

\subsection{Evolution}

The  general implementation of the memetic algorithm is described by Algorithm \ref{pseudcode-mem-evol}.\footnote{The notation is inspired by a pseudocode given by \citeN{merz_memetic_2012}.} 
The genetic operators of crossover, mutation, and selection (described below) are applied to each generation of communities. Subgraphs generated by crossover and mutation are adapted by a local search. If the starting subgraph is not connected we replace it by its main component. Evolution is terminated when no better best community is found for many generations.

\begin{algorithm}[!b]
\caption{Pseudocode of memetic evolution for one adapted seed}
\label{pseudcode-mem-evol}
 \begin{algorithmic}
 \STATE    
\STATE  \textbf{initialise} population \textbf{P} by
\textbf{mutating} the adapted seed with high variance several times and \textbf{adapting} mutants
\WHILE {the best community is not too old}
\STATE  \textbf{mutate} the best community with low variance and \textbf{adapt} the mutants
\IF {an adapted mutant is new and its cost is lower than highest cost}
\STATE  \textbf{add} it to population \textbf{P} 
\ENDIF 
\STATE \textbf{cross} the best community with some randomly chosen communities and \textbf{adapt} the offspring
\IF {adapted offspring is new and its cost is lower than highest cost}
\STATE  \textbf{add} it to population \textbf{P} 
\ENDIF 
\STATE  \textbf{select} the best communities so that the population size remains constant 
\IF {there is no better best community for some generations and innovation rate is low} 
\STATE  \textbf{renew} the population by mutating the best community with high variance and \textbf{adapt} mutants 
 \STATE  \textbf{select} the best communities so that the population size remains constant 
 \ENDIF 
   \ENDWHILE
 \STATE
 \end{algorithmic}
\end{algorithm}

\subsection{Genetic operators}

\paragraph{Mutation:}

We mutate a community 
with mutation variance $v < 1$ by changing  maximally a 
proportion $v$ of its links or nodes. 
In node-wise memetic evolutions we randomly exclude boundary nodes and then include the same number of neighbouring nodes. In link-wise memetics we experiment with two other mutation operators: we only exclude \textit{or} include links and concentrate  changes around one randomly chosen boundary node. (Details can be found in Appendix of Part~2.)

\paragraph{Crossover:}

From two parent subgraphs we construct two new individuals by taking intersection and union of the subgraphs as starting points for adaptive local searches. 
Of course, it has no effect to cross such parents where one of them is part of the other one. 
Normally, evolutionary algorithms include some randomness in the crossover, which in our case would mean to enlarge the intersection by some nodes or links from the union. 
In contrast, our crossing procedure is deterministic because the boundary of the union of two good communities should also be not too bad. The same holds for the intersection.
Deterministic crossover should be (and is) done only once with the same parents. 
The only random element of our crossover is the random selection of parents.

\paragraph{Selection:}
From the old population and the results of mutations and crossovers we select the communities with lowest $\Psi$-values, keeping the population size constant. A new best  community is only included if it is inside the 
minimal range of the  best community of the original population.
Disregarding the best communities
outside the minimal  range 
assures that we do not lose communities which can have a range above 
the minimum given by the resolution limit we apply.
Deselected communities can be used as seeds for other memetic searches. 

\paragraph{Renewal:}
Renewal means to mutate the best community with high variance several times, to adapt the mutants, and to apply a usual selection procedure described above.  

\section{Concluding remarks}

In the  forthcoming Part 2 of our paper 
we discuss test results.

\section*{Acknowledgement}
This work is part of a project in which we develop methods for  measuring the diversity of research. The project was funded by the German Ministry for  Education and Research (BMBF). We would like to thank the members of the project's advisory group\footnote{\url{http://141.20.126.172/~div}} and also all developers of \textbf{R}.\footnote{\url{http://www.r-project.org}} 
We thank Alexander Struck for assisting our work and for discussions about the new cost function.  We discussed the issue of link similarity in citation networks with Rob Koopman. 

\newpage
\section*{Appendix}
\begin{appendix}

\section{Connectivity measures}
\label{app:sigma}

In this section we derive the connectivity measures for link 
sets
$\sigma(L)$ and $k_\mathrm{in}(L)$ from analogue measures in the line graph.
We closely follow the arguments given in our earlier paper \cite{havemann2012evaluating}.

We here use $i, j = 1, \ldots, n$ to denote nodes and $k, l = 1, \ldots, m$ for links.
With $C(L)$ we name the set of nodes attached to links in  
the subgraph induced by link set
$L$. If a link $k$ belongs to 
$L$ its membership $\mu_k(L)=1$ and zero otherwise.

To construct a network's line graph we first define an auxiliary bipartite graph obtained by putting a node on each link of the original network. The affiliation matrix $B$ of the bipartite graph---also called its incidence matrix---has a row for each of the $n$ original nodes and a column for each of the $m$ original links. Each link column contains only two non-zero elements, namely the elements in the rows of the nodes $i$ and $j$ connected by the link. We can project the bipartite graph back onto the original network with the product $BB^\mathrm{T}$ which equals its adjacency matrix $A$  (except for the main diagonal). 

We obtain the network's line graph by the opposite projection $B^\mathrm{T}B$ of the bipartite graph. \citeN{evans2009line} underline, that in all cases of practical intererest the line graph contains the same amount of information as the original network. Knowing $B^\mathrm{T}B$ we can almost ever calculate $BB^\mathrm{T}$ and thus also the network's adjacency matrix $A$.

Because each node of the original network is represented as a clique in the line graph \citeN{evans2009line} weighted the edges of the line graph with the inverse degree $1/k_i$ of the node $i$ in the original network. They define the line graph's adjacency matrix as
\begin{equation}
E_{kl} = \sum_{i=1}^n \frac{B_{ik}B_{il}}{k_i}.                                               
\label{eq:E}
\end{equation}

Weighting the line graph's edges with the inverse degrees of  nodes in the original network is equivalent to an Euclidean normalisation of the nodes' vectors in the affiliation matrix $B$ of the auxiliary bipartite graph. This becomes clear if we factorise the terms of the sum  in equation \ref{eq:E}:  
\begin{equation}
E_{kl} = \sum_{i=1}^n \frac{B_{ik}}{\sqrt{k_i}}\frac{B_{il}}{\sqrt{k_i}}.
\label{eq:factorise}
\end{equation}
Then we can shortly write $E=D^\mathrm{T}D$ with $D_{ik}=B_{ik}/\sqrt{k_i}$ and verify the Euclidean normalisation of the $n$ row vectors of $D$ (for unweighted networks for which we have $B_{ik}^2=B_{ik}$):
\begin{equation}
\sum_{k=1}^m D_{ik}^2=  \sum_{k=1}^m\frac{B_{ik}^2}{k_i} =  \frac{1}{k_i}\sum_{k=1}^m B_{ik}=1.
\label{eq:L2}
\end{equation}

On the other hand, the projection of the normalised bipartite graph described by affiliation matrix $D$ back on a network of the original nodes is given by $DD^\mathrm{T}$. An element of adjacency matrix $DD^\mathrm{T}$ is given by 
\begin{equation}
\sum_{k=1}^m D_{ik}D_{jk} = \sum_{k=1}^m\frac{B_{ik}B_{jk}}{\sqrt{k_ik_j}} =  \frac{A_{ij}}{\sqrt{k_ik_j}}.
\label{eq:back-projection}
\end{equation}
Thus, Euclidean normalisation of $B$'s row vectors is equivalent to weighting each link in the original (unweighted) network with the geometric mean of its nodes' inverse degrees.  The weighted graph described by adjacency matrix $E$ is not the line graph of the unweighted network  described by adjacency matrix $A$ but of the network weighted according to equation \ref{eq:back-projection}. 
It depends on the real relations we model with the network whether this is a realistic weighting.

Now we calculate internal connectivity $\tau(L)$ as 
the sum of internal degrees of vertices in the line graph: 
\begin{equation}
\begin{split}
\tau(L)&=\sum_{k, l = 1}^m \mu_k(L)E_{kl}\mu_l(L)\\
&=\sum_{k, l = 1}^m \mu_k(L)  \sum_{i=1}^n \frac{B_{ik}B_{il}}{k_i}  \mu_l(L).
\end{split}
\end{equation} 
In the same way, we can calculate external connectivity $\sigma(L)$ as the sum of external degrees in the line graph:
\begin{equation}
\begin{split}
\sigma(L)&=\sum_{k, l = 1}^m \mu_k(L)E_{kl}(1-\mu_l(L)).\\
&=\sum_{k, l = 1}^m \mu_k(L)\sum_{i=1}^n \frac{B_{ik}B_{il}}{k_i}(1-\mu_l(L)).
\end{split}
\end{equation}
Now we use the relations
$$\sum_{k=1}^m \mu_k(L) B_{ik} = k_i^\mathrm{in}(L)$$
and
$$\sum_{l=1}^m (1-\mu_l(L))B_{il}= k_i^\mathrm{out}(L),$$
which directly follow from the definition of the incidence matrix $B$. Thus, we get 
$$\tau(L)=\sum_{i=1}^n \frac{(k_i^\mathrm{in}(L))^2}{k_i}$$
and
$$\sigma(L)=\sum_{i=1}^n \frac{k_i^\mathrm{in}(L)k_i^\mathrm{out}(L)}{k_i}.$$
From this we easily derive total connectivity of a link-induced subgraph as the sum
$$ \tau(L) + \sigma(L)= \sum_{i=1}^n k_i^\mathrm{in}(L)=k_\mathrm{in}(L). $$

\section[Cost-landscape of the bow-tie graph]{Cost-landscape\\of the bow-tie graph}
\label{sec:bowtie}
\begin{figure}[!t] 
\begin{center}  \includegraphics[width=2in]{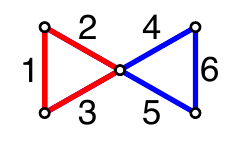} 
\end{center} 
\caption{Bow-tie graph with numbered links} \label{bow-tie-app} 
\end{figure}

For the bow-tie graph we expect two link communities, namely the triangle $\{1, 2, 3\}$ and its complement $\{4, 5, 6\}$, cf.\ Figure \ref{bow-tie-app} and \citeN{evans2009line}.
To describe the $2^m$ different possible subgraphs it is advantageous to make use of the spherical topology of any landscape of subgraphs.
Indeed, the cost-function landscape of a graph's subgraphs can be seen as the surface of a globe
\begin{itemize}
\item 
 with the whole and the empty graph at the poles,
\item with all possible subgraphs of the same size on each circle of latitude, and
\item  with complementary subgraphs situated at antipodes.
\end{itemize}
The neighbours of a place in the landscape can be reached by adding an element to the set of nodes (for node-induced subgraphs) or of links (for link-induced subgraphs), respectively, or by deleting an element from this set.
That means,
there are no direct relations between places on the same circle of latitude.
Steps (adding or removing nodes or links) are moves between neighbouring circles of latitude.

We define  the north pole as corresponding to the empty subgraph and the south pole as corresponding to the whole graph.
  The $\Psi$-globe of the bow-tie graph has five circles of latitude corresponding to six subgraphs with one link, 15 with two, 20 with three, 15 with four, and six with five links, respectively.

For the empty graph at the north pole $\sigma = 0$  and $\Psi=1$ (by definition). 
The six single links as the smallest real subgraphs are located at the highest circle of latitude.  
The two outer links 1 and 6 have $\sigma = 1\cdot 1/2 + 1\cdot1/2 = 1$ and $\Psi=0.6$,
the four inner links have $\sigma = 1\cdot 1/2 + 1\cdot 3/4 = 5/4$  and $\Psi=0.75$.

 There are ten connected and five unconnected subgraphs with two links:
  \begin{itemize}
\item 
four connected subgraphs with one outer link and one inner link (e.g.\ link set $\{1, 2\}$) resulting in $\sigma = 1\cdot 1/2 + 1 \cdot3/4 = 5/4$  and $\Psi \approx 0.469$,
\item six connected subgraphs with two inner links (e.g.\ link set $\{2, 3\}$) and $\sigma = 1\cdot 1/2 + 2\cdot 2/4 + 1 \cdot1/2 = 2$  and $\Psi=0.75$,
\item  four unconnected subgraphs with one outer and one inner link (e.g.\ link set $\{1, 4\}$) and $\sigma = 9/4$  and $\Psi\approx 0.844$,
 \item  one unconnected subgraph  with two outer links ($\{1, 6\}$) and $\sigma = 2$  and $\Psi=0.75 $.
\end{itemize}
  
On the equator of the $\Psi$-globe there are 20 triples of links which can be classified into four types:
 \begin{itemize}
\item 
the triangle $ \{1, 2, 3\}$ and its complement $\{4, 5, 6\}$ with $\sigma = 2 \cdot2/4 = 1$  and $\Psi=1/3$,
\item four triples of inner links (e.g.\ link set $\{2, 3, 4\}$) and their unconnected complements (e.g.\ link set $\{1, 5, 6\}$) with $\sigma =  3/2 + 3/4 = 9/4$  and $\Psi=0.75 $,
\item eight subgraphs with one of the two outer links, one of the two attached inner links and one of the two inner links not attached to the outer link
(e.g.\ link set $\{1, 2, 4\}$): they all have $\sigma = 1 \cdot1/2 + 2 \cdot2/4 + 1\cdot 1/2 = 2$  and $\Psi=2/3 $,
 \item  the unconnected triple with one outer and two inner links (set $\{1, 4, 5\}$) and its unconnected complement (set $\{2, 3, 6\}$) with $\sigma = 3$  and $\Psi=1$.   \end{itemize}

On the two circles of latitude on the southern hemisphere we find the complements of the subgraphs on the northern hemisphere with the same $\Psi$-values. 
Next to the equator we find 13 connected and two unconnected subgraphs with four links each: 
  \begin{itemize}
\item 
two unconnected quadruples with one triangle and the second outer link (e.g.\ link set $\{1, 2, 3, 6\}$) which have $\sigma=2$  and $\Psi=0.75$,
\item  the central star with all four inner links $\{2, 3, 4, 5\}$ with $\sigma$ = 4/2 = 2  and $\Psi=0.75$,
\item  four subgraphs containing one of the two triangles plus one of the two inner links (e.g.\ link set $\{1, 2, 3, 5\}$) which all have $\sigma = 1\cdot 3/4 + 1\cdot 1/2 = 5/4$  and $\Psi\approx 0.469$,
\item  the four subgraphs with both outer links and two inner links connecting them (e.g.\ link set $\{1, 2, 5, 6\}$) with $\sigma = 2/2 + 4/4 = 2$  and $\Psi=0.75$,
\item  the four subgraphs with one outer link and three inner links (one of them attached to the outer link, e.g.\ link set $\{1, 2, 4, 5\}$) with $\sigma = 3/2 + 3/4 = 9/4$  and $\Psi\approx 0.844$.
 \end{itemize}
All complements of the six single links containing the five other links are connected and have the same $\Psi$-values as their single-link complements  (cf. above).
The full graph at the south pole with $\sigma = 0$  and $\Psi=1$ is connected. 
The  $\Psi$-landscape of links has two local minima:
  the two triangles have a locally and globally minimal $\Psi=1/3$.  
 There are no other local minima. 
  Thus, we obtain the pair of complementary triangles as the only solution.

\end{appendix}

\bibliography{informetrics}

\begin{thebibliography}{}

\bibitem[\protect\citeauthoryear{Ahn, Bagrow, and Lehmann}{Ahn
  et~al.}{2010}]{ahn2010link}
Ahn, Y., J.~Bagrow, and S.~Lehmann (2010).
\newblock Link communities reveal multiscale complexity in networks.
\newblock {\em Nature\/}~{\em 466\/}(7307), 761--764.

\bibitem[\protect\citeauthoryear{Amelio and Pizzuti}{Amelio and
  Pizzuti}{2014}]{amelio_overlapping_2014}
Amelio, A. and C.~Pizzuti (2014).
\newblock Overlapping {Community} {Discovery} {Methods}: {A} {Survey}.
\newblock {\em Social Networks: Analysis and Case Studies\/}, 105.
\newblock \url{http://arxiv.org/abs/1411.3935}.

\bibitem[\protect\citeauthoryear{Clauset}{Clauset}{2005}]{clauset2005flc}
Clauset, A. (2005).
\newblock {Finding local community structure in networks}.
\newblock {\em Physical Review E\/}~{\em 72\/}(2), 26132.

\bibitem[\protect\citeauthoryear{Evans and Lambiotte}{Evans and
  Lambiotte}{2009}]{evans2009line}
Evans, T. and R.~Lambiotte (2009).
\newblock {Line graphs, link partitions, and overlapping communities}.
\newblock {\em Physical Review E\/}~{\em 80\/}(1), 16105.

\bibitem[\protect\citeauthoryear{Fortunato}{Fortunato}{2010}]{fortunato2010community}
Fortunato, S. (2010).
\newblock Community detection in graphs.
\newblock {\em Physics Reports\/}~{\em 486\/}(3-5), 75--174.

\bibitem[\protect\citeauthoryear{Gach and Hao}{Gach and
  Hao}{2012}]{gach_memetic_2012}
Gach, O. and J.-K. Hao (2012, January).
\newblock A memetic algorithm for community detection in complex networks.
\newblock In C.~A.~C. Coello, V.~Cutello, K.~Deb, S.~Forrest, G.~Nicosia, and
  M.~Pavone (Eds.), {\em Parallel Problem Solving from Nature - {PPSN} {XII}},
  Number 7492 in Lecture Notes in Computer Science, pp.\  327--336. Springer
  Berlin Heidelberg.

\bibitem[\protect\citeauthoryear{Gong, Fu, Jiao, and Du}{Gong
  et~al.}{2011}]{gong_memetic_2011}
Gong, M., B.~Fu, L.~Jiao, and H.~Du (2011, November).
\newblock Memetic algorithm for community detection in networks.
\newblock {\em Physical Review E\/}~{\em 84\/}(5), 056101.

\bibitem[\protect\citeauthoryear{Havemann, Gl{\"a}ser, Heinz, and
  Struck}{Havemann et~al.}{2012}]{havemann2012evaluating}
Havemann, F., J.~Gl{\"a}ser, M.~Heinz, and A.~Struck (2012).
\newblock Evaluating overlapping communities with the conductance of their
  boundary nodes.
\newblock {\em arXiv preprint arXiv:1206.3992\/}.

\bibitem[\protect\citeauthoryear{Havemann, Heinz, Struck, and
  Gl{\"a}ser}{Havemann et~al.}{2011}]{havemann2011identification}
Havemann, F., M.~Heinz, A.~Struck, and J.~Gl{\"a}ser (2011).
\newblock {Identification of Overlapping Communities and their Hierarchy by
  Locally Calculating Community-Changing Resolution Levels}.
\newblock {\em Journal of Statistical Mechanics: Theory and Experiment\/}~{\em
  2011}, P01023.
\newblock doi: 10.1088/1742-5468/2011/01/P01023, Arxiv preprint
  arXiv:1008.1004.

\bibitem[\protect\citeauthoryear{Hric, Darst, and Fortunato}{Hric
  et~al.}{2014}]{hric_community_2014}
Hric, D., R.~K. Darst, and S.~Fortunato (2014, December).
\newblock Community detection in networks: Structural communities versus ground
  truth.
\newblock {\em Physical Review E\/}~{\em 90\/}(6), 062805.

\bibitem[\protect\citeauthoryear{Lancichinetti, Fortunato, and
  Kertesz}{Lancichinetti et~al.}{2009}]{lancichinetti2009detecting}
Lancichinetti, A., S.~Fortunato, and J.~Kertesz (2009).
\newblock {Detecting the overlapping and hierarchical community structure in
  complex networks}.
\newblock {\em New Journal of Physics\/}~{\em 11}, 033015.
\newblock
  arXiv:\href{http://arxiv.org/abs/arXiv:0802.1218}{physics.soc-ph/0802.1218}.

\bibitem[\protect\citeauthoryear{Li, Zhang, Wang, Liu, and Zhang}{Li
  et~al.}{2013}]{li_discovering_2013}
Li, Z., X.-S. Zhang, R.-S. Wang, H.~Liu, and S.~Zhang (2013, December).
\newblock Discovering link communities in complex networks by an integer
  programming model and a genetic algorithm.
\newblock {\em {PLoS} {ONE}\/}~{\em 8\/}(12), e83739.

\bibitem[\protect\citeauthoryear{Ma, Gong, Liu, Cai, and Jiao}{Ma
  et~al.}{2014}]{ma_multi-level_2014}
Ma, L., M.~Gong, J.~Liu, Q.~Cai, and L.~Jiao (2014, June).
\newblock Multi-level learning based memetic algorithm for community detection.
\newblock {\em Applied Soft Computing\/}~{\em 19}, 121--133.

\bibitem[\protect\citeauthoryear{Merz}{Merz}{2012}]{merz_memetic_2012}
Merz, P. (2012, January).
\newblock Memetic algorithms and fitness landscapes in combinatorial
  optimization.
\newblock In F.~Neri, C.~Cotta, and P.~Moscato (Eds.), {\em Handbook of Memetic
  Algorithms}, Number 379 in Studies in Computational Intelligence, pp.\
  95--119. Springer Berlin Heidelberg.

\bibitem[\protect\citeauthoryear{Neri, Cotta, and Moscato}{Neri
  et~al.}{2012}]{neri2012handbook}
Neri, F., C.~Cotta, and P.~Moscato (Eds.) (2012).
\newblock {\em {Handbook of Memetic Algorithms}}, Volume 379 of {\em Studies in
  Computational Intelligence}.
\newblock Springer, Berlin.

\bibitem[\protect\citeauthoryear{Pan, Wang, and Xie}{Pan
  et~al.}{2012}]{pan_link_2012}
Pan, L., C.~Wang, and J.~Xie (2012).
\newblock Link communities detection via local approach.
\newblock In T.~Li, H.~Nguyen, G.~Wang, J.~Grzymala-Busse, R.~Janicki,
  A.~Hassanien, and H.~Yu (Eds.), {\em Rough Sets and Knowledge Technology},
  Volume 7414 of {\em Lecture Notes in Computer Science}, pp.\  282--291.
  Springer Berlin / Heidelberg.

\bibitem[\protect\citeauthoryear{Pan, Wang, Xie, and Liu}{Pan
  et~al.}{2011}]{pan_detecting_2011}
Pan, L., C.~Wang, J.~Xie, and M.~Liu (2011).
\newblock Detecting link communities based on local approach.
\newblock In {\em 2012 {IEEE} 24th International Conference on Tools with
  Artificial Intelligence}, Volume~0, Los Alamitos, {CA}, {USA}, pp.\
  884--886. {IEEE} Computer Society.

\bibitem[\protect\citeauthoryear{Pizzuti}{Pizzuti}{2009}]{pizzuti_overlapped_2009}
Pizzuti, C. (2009).
\newblock Overlapped community detection in complex networks.
\newblock In {\em {GECCO}}, Volume~9, pp.\  859{\textendash}866.

\bibitem[\protect\citeauthoryear{Pizzuti}{Pizzuti}{2012}]{pizzuti_boosting_2012}
Pizzuti, C. (2012).
\newblock Boosting the detection of modular community structure with genetic
  algorithms and local search.
\newblock In {\em Proceedings of the 27th Annual {ACM} Symposium on Applied
  Computing}, pp.\  226{\textendash}231.

\bibitem[\protect\citeauthoryear{Shi, Cai, Fu, Dong, and Wu}{Shi
  et~al.}{2013}]{shi_link_2013}
Shi, C., Y.~Cai, D.~Fu, Y.~Dong, and B.~Wu (2013, September).
\newblock A link clustering based overlapping community detection algorithm.
\newblock {\em Data \& Knowledge Engineering\/}~{\em 87}, 394--404.

\bibitem[\protect\citeauthoryear{Vitela and Casta{\~n}os}{Vitela and
  Casta{\~n}os}{2012}]{vitela_sequential_2012}
Vitela, J.~E. and O.~Casta{\~n}os (2012, May).
\newblock A sequential niching memetic algorithm for continuous multimodal
  function optimization.
\newblock {\em Applied Mathematics and Computation\/}~{\em 218\/}(17),
  8242--8259.

\bibitem[\protect\citeauthoryear{Wei and Cheng}{Wei and
  Cheng}{1989}]{wei_towards_1989}
Wei, Y.-C. and C.-K. Cheng (1989, November).
\newblock Towards efficient hierarchical designs by ratio cut partitioning.
\newblock In {\em {IEEE} International Conference on Computer-Aided Design,
  1989. {ICCAD}-89. Digest of Technical Papers}, pp.\  298--301.

\bibitem[\protect\citeauthoryear{Xie, Kelley, and Szymanski}{Xie
  et~al.}{2013}]{Xie:2013:OCD}
Xie, J., S.~Kelley, and B.~K. Szymanski (2013, August).
\newblock Overlapping community detection in networks: The state-of-the-art and
  comparative study.
\newblock {\em ACM Comput. Surv.\/}~{\em 45\/}(4), 43:1--43:35.

\bibitem[\protect\citeauthoryear{Yang and Leskovec}{Yang and
  Leskovec}{2012}]{yang2012defining}
Yang, J. and J.~Leskovec (2012).
\newblock Defining and evaluating network communities based on ground-truth.
\newblock In {\em Proceedings of the ACM SIGKDD Workshop on Mining Data
  Semantics}, pp.\ ~3. ACM.

\end{thebibliography}
\bibliographystyle{chicago}

\end{document}